\documentclass[runningheads]{llncs}
\usepackage{graphicx}
\usepackage{comment}
\usepackage{makecell}
\usepackage{color}
\usepackage{pifont}
\usepackage{url}
\usepackage{hyperref}
\usepackage{float}

\newcommand{\cmark}{\ding{51}}%
\newcommand{\xmark}{\ding{55}}%

\newcolumntype{C}[1]{>{\centering\arraybackslash}p{#1}}

\begin{document}
	\title{A Spark is Enough in a Straw World: a Study of Websites Password Management in the Wild}
	\titlerunning{A Spark is Enough in a Straw World}
	%
	\author{Simone Raponi \and Roberto Di Pietro}
	\authorrunning{Simone Raponi and Roberto Di Pietro}
	%
	
	\institute{Hamad Bin Khalifa University,\\College of Science and Engineering,\\Doha, Qatar \\
		\email{\{sraponi,rdipietro\}@hbku.edu.qa}\\}

	\maketitle
	\begin{abstract}


		The widespread usage of password authentication in online websites leads to an ever-increasing concern, especially when considering  the possibility for an attacker to recover the user password by leveraging the loopholes in the password recovery mechanisms. Indeed, if a website adopts a poor  password management system,  
		this choice makes useless
		even the most robust password chosen by its users. 
		In this paper, we first provide a survey of currently adopted password recovery mechanisms. Later, we model an attacker with different capabilities and we show how current password recovery mechanisms can be exploited in our attacker model. 
		Then, we provide a thorough analysis of the password management of some of the Alexa's top 200 websites in different countries, including England, France, Germany, Spain and Italy. Of these 1,000 websites, 722 do not require authentication---and hence are excluded by our study---, while out of the remaining 278 we focused on 174---since 104 demanded a complex registration procedure. Of these 174,  almost 25\% of the  them have critical vulnerabilities, while 44\% have some form of vulnerability. Finally, we propose some effective countermeasures and we point out that, by considering the entry into force of the General Data Protection Regulation (GDPR) in May, 2018, most of websites are not compliant with the legislation and may incur in heavy fines. This study, other than being important on its own since it highlights some severe current vulnerabilities and proposes corresponding remedies, has the potential to also have a relevant impact on the EU industrial ecosystem.

	\end{abstract}
	\section{Introduction}	
	
	The countless attacks on websites in recent years have underlined once again that security is often considered a feature, rather than a necessity~\cite{website_security_statistics_report}. The victims of such attacks are the users that, unaware of the management of the provided confidential information, will find their identity compromised on the web. In fact, even if a user would adopt all known good practices to choose a really strong password to enforce access control, a single website that stores information in an insecure way would be enough to compromise confidentiality at least. Examples are the data breaches suffered by both the professional networking site LinkedIn~\cite{linkedin_data_breach} and the Internet service company Yahoo~\cite{yahoo_data_breach}.
	LinkedIn suffered a major breach in June 2012, when 6.5 million encrypted passwords were posted on a Russian website. Things got much worse in  May 2016, when 117 million LinkedIn credentials (i.e., combination of e-mails and passwords) were posted for sale on the Dark Web~\cite{linkedin_dark_web}.
	Although LinkedIn was a well-known and well-established website, poor cryptography techniques were implemented~\cite{kamp2012linkedin}, by making easier for hackers to decrypt users' passwords. The attack targeting Yahoo was the biggest one in history, which led to the breach of 3 billion passwords. The hack began with a spear-phishing e-mail sent to a Yahoo employee, that eventually led to the acquisition of the entire users database by the attacker (containing names, phone numbers, password challenge questions and answers, password recovery e-mails and a cryptographic value unique to each account)~\cite{yahoo_russian_hack}.
	If it were true that history taught, we should have solved (or at least mitigated) problems of this kind, but reality is different. In fact, in this work we point out that almost 44\% of the top Alexa's websites we considered (out of the top 200 of respectively England, France, Germany, Italy, and Spain, shown in \autoref{tab:analyzed_websites}) do store users password in a form that can be easily exploited. While the impact of these findings are relevant on their own, their consequences for companies are magnified when taking into account the GDPR---the poor security measures adopted by the web sites are a clear violation of. \newline
	The General Data Protection Regulation (GDPR)~\cite{gdpr_portal} leads to an important change in the vision of both data privacy and data security. Born as an evolution of the previous Data Protection Directive, adopted in 1995, it becomes enforceable from May 25th, 2018. With its entry into force, all the websites, agencies, enterprises, organizations, that make use of the personal data of the users have to guarantee their protection both by design and default in any operation.
	In case of non-compliance, these entities are subject to very heavy penalties ranging from 10-20 million euros to 2-4\% of the annual worldwide turnover of the previous financial year, unsustainable by most organizations.
	\paragraph{Contributions.}
	In this paper we first provide a survey of both users authentication mechanisms implemented by websites and related password recovery mechanisms.
	Then we model a realistic attacker with different capabilities, respectively Mail Service Provider attacker, Web Server Intruder attacker, Client Intruder attacker, and Sniffing attacker. Later we provide a thorough analysis for users password management Alexa's top 200 websites of the five aforementioned European countries.
	Then, we study in detail which information could be obtained by our modeled attacker and we show how she can break the access control mechanisms. \\
	Results are striking; of the 174 analyzed websites (see \autoref{tab:analyzed_websites}) almost 25\% of the websites do have from poor to very poor password management, and an overall of 43.68\% are vulnerable to at least one of the presented attacks---note that all the attacks happen because of the non-compliance of the websites to the GDPR prescriptions, hence having the corresponding organizations being subject to the cited fines. Finally we present some effective countermeasures that do match the GDPR requirements and that would prevent such attacks to occur. \newline \newline
	\noindent \textbf{Road-map:} In Section 2, we report on the related work in the literature. In Section 3, we provide a technical background of both user authentication on websites and the password recovery mechanisms. In Section 4, we define our attacker model. In Section 5, we describe the methodology we adopted and we present the results of the analysis. In Section 6 we propose countermeasures to mitigate the discovered vulnerabilities, while in Section 7 we report some concluding remarks.
	
	\section{Related Work}
	LinkedIn data breaches in June 2012~\cite{linkedin_data_breach} and Yahoo data breaches in August 2013~\cite{yahoo_data_breach}, made respectively 6.5 million and 3 billion users accounts compromised, but these compromises are only the tip of the iceberg. However, these attacks have not affected the popularity of passwords. In fact, passwords remain the most widespread authentication mechanism on the web. The use of a password introduces a secret that is shared by only the authenticator (the website), and the user wishing to be authenticated. \newline
	From the moment they were adopted, a number of scientific articles were published with the aim of highlighting their weaknesses and vulnerabilities. In~\cite{florencio2007large} the authors analyzed half a million Windows Live users' passwords, pointing out that a user has 6.5 passwords shared across 3.9 different sites on average. Furthermore, each user is the owner of about 25 accounts and types an average of 8 passwords per day. Most of the picked passwords are extremely weak, in fact, if not forced, users choose passwords composed by solely lowercase letters. Dell'Amico et al., in~\cite{dell2010password} focused on the empirical study of real-world passwords. They implemented and used several state-of-the-art techniques for password guessing to analyze the password  strength of Internet application. They found that users put relatively little effort in choosing their password when compared to the choice of their usernames.
	The human component plays a fundamental role in both the security and the robustness of the authentication mechanisms. Indeed, even the most advanced system would be compromised if users pay little attention to their password choice.  \newline
	Passwords authentication mechanisms will still be used for years, as ``something you know'' mechanisms are extremely less expensive (but also less secure) than both ``something you have'' and ``something you are''---these latter ones being prone also to false positive and false negative. By considering this, it is of fundamental importance to guarantee both secure access and secure storage, as well as secure mechanism to retrieve the password in case of forgetfulness or theft.
	In~\cite{furnell2007assessment}, the authors presented an assessment of password practices on 10 popular websites, including Facebook, Amazon, Yahoo, Google, and YouTube. They examined password selection, the restrictions enforced on password choice, and the recovery/reset of the password if forgotten. They pointed out that no website provides adequate coverage of all the criteria taken into account. The result is worrying, as the websites analyzed are among the most visited on the web and therefore (should be) the safest.
	\newline
	When we lose or forget our passwords we generally use our e-mail address to retrieve them, assuming that we are the only ones who have the access. S.L. Garfinkel, in~\cite{garfinkel2003email}, wondered if an e-mail-Based Identification and Authentication (EBIA) method could substitute the Public Key Infrastructure (PKI). The EBIA technique considers an e-mail address as a universal identifier and the ability to receive an e-mail at that address as a kind of authenticator. Among the many limitations, the author pointed out two many vulnerabilities: EBIA security strongly depends on the security of e-mail servers and password; e-mail content is accessible to server operators (without encryption, system managers can intercept, read, and make copies of e-mail messages intended for end users). Although the paper is dated back 2003, EBIA is the main used secondary (also known as emergency) authentication method nowadays, with the same vulnerabilities (and a few additional ones) left unresolved. One possible countermeasure would be to use secure e-mail services, such as ProtonMail~\cite{protonmail}. ProtonMail is an open-source end-to-end encrypted e-mail service founded in 2014 at the CERN. This service allows to protect the user's privacy and anonymity by not logging IP addresses which can be linked to the account, furthermore the end-to-end encryption makes unreadable the e-mail content even to the e-mail provider. \newline
	In recent years, various alternatives were proposed to manage the secondary authentication mechanism in a safer way. After ``something you have'', ``something you are'', and ``something you know'', in~\cite{brainard2006fourth} the authors explored a fourth-factor authentication: ``somebody you know''. They focused in a process called vouching, a peer-level authentication in which a user (called helper), leverages her primary authenticator to assist a second user (called asker) to perform secondary authentication. They designed a prototype vouching system for SecurID, a hardware authentication token, to allow an helper to grant temporary access privileges to an asker who has lost the ability to use her own. Although interesting, this method requires a primary authentication mechanism based on token, hence ``something you have'', which is not implemented by most web sites. In~\cite{schechter2009s} Schechter et al. exploited the social-authentication to let users who have forgotten their passwords to regain access to their account. The proposed system employs trustees, users previously appointed by account holders to verify their identity. To get the access into her account, the account holder contacts their trustees in person or by phone, so that their trustees may recognize her by either her appearance or her voice. Once the recognition has occurred, the trustees provide the account holder with an account recovery code, that will be necessary to authenticate her in the system. This mechanism is safer with respect to the ones currently adopted but may have usability problems. Trustees may not be available at the time of request, making it impossible to the requesting user to recover her account. \newline
	The secondary authentication mechanisms are discussed and analyzed in~\cite{reeder2011password}. In this work the authors considered four key criteria (i.e., reliability, security, authentication, and setup efficiency) to evaluate several secondary authentication mechanisms, such as security questions, printed shared secrets, previously used passwords, e-mail-based verification, phones or other services, trustees, and in-person proofing. Although they provided a thorough analysis of these mechanisms, no attackers were modeled and no experimental website security analysis was performed.

	\section{Technical Background}
	In this section, a technical background of both user authentication mechanisms on websites and password recovery mechanisms adopted by websites are provided.
	

	\subsection{User authentication on websites}
	With user authentication on websites we refer to the process in which the credentials provided by the user are compared to those stored either in websites database or in a cloud server. If the credentials match, the authentication process is completed and the requesting user is granted authorization to access. According to both the website type and restrictions, there are a few user authentication methods~\cite{furnell2007comparison}. A combination of these methods leads to a more accurate identification of the user:
	\begin{itemize}
		\item \textbf{1FA}: the 1-Factor-Authentication (1FA) method requires only one factor to authenticate a user. Usually it takes into account ``something the user knows'' (e.g., a password or a PIN code). 1FA is the authentication mechanism most commonly adopted by current websites;
		\item \textbf{2FA}: the 2-Factor-Authentication (2FA) method requires two factors to authenticate a user. It takes into account both ``something the user knows'' and either ``something the user has'' (e.g., physical token or a smart card) or ``something the user is'' (e.g., fingerprints, retinal or iris scans, voice recognition, hand geometry). This mechanism is adopted in most of the sensitive websites (e.g., bank websites) and is optional in others (e.g., Gmail~\cite{google_2fa}, ProtonMail~\cite{protonmail_2fa});
		\item \textbf{3FA}: the 3-Factor-Authentication (3FA) method requires a user being authenticated with ``something she knows'' as well as with both ``something she has'' and ``something she is''. This mechanism has not been adopted by the websites, probably due to the infrastructure costs that would derive from it.
		
	\end{itemize}

	\subsection{Password recovery}
	Password recovery is a mechanism implemented by websites that allows to recover the user's secret password in case the password is lost or forgotten. During the years, numerous mechanisms have been proposed to recover the password, the most frequently adopted are reported below.
	
	\subsubsection{Security Questions.}
	When the account holder loses or forgets her password, the password recovery mechanism starts up and provides the user with security questions. This mechanism is based on the assumption that only the account holder is able to answer correctly. Some websites make use of a set of pre-packaged standard questions, while others allow the users to choose their own. In each case, several work demonstrated both limits and vulnerabilities of this mechanism. Using a set of pre-packaged password is quite insecure in the era of the information; answering questions like ``what is the name of your primary school?'' or ``what is your favorite movie?'' becomes trivial by having access to all the personal information that the user publishes on the various social networks. On the other hand, users could choose their own customized security questions. Even in this case, several work~\cite{just2009personal},\cite{reeder2011password} demonstrated how weak the mechanism is. As they mention, users have to select questions that are memorable, not researchable on-line, reasonably unpopular with other users, and unknown by any untrusted acquaintances.
	
	\subsubsection{Previously used passwords.}
	Websites, as a password recovery mechanism, could ask the user to enter one (or a set of) previously used password(s). However, users tend to use a limited number of passwords for the web services they access, this phenomenon leading to the unavoidable use of the same password for one or more services. Given that, it should be easy for users to remember one of the password previously used---with high probability it will be a password they are currently using for another web service. Although with both some limitations (e.g., the user may not remember any of the previous passwords) and many security problems (e.g., the attacker may know some of the previous passwords of the victim that could be in use to access other services on different platforms) this mechanism could be useful and less dangerous than the ones currently implemented. It is worth noting that this method cannot be applied if the password has been lost or stolen for the first time.
	
	\subsubsection{E-mail-based authentication.}
	The most common password recovery mechanism is e-mail-based authentication, which relies on the assumption that only the account holder will be able to access a secret sent to her e-mail account~\cite{garfinkel2003email}\cite{reeder2011password}. As we will discuss in Section 5, websites can provide the requesting user with her credential information in several ways: 
	\begin{itemize}
		\item sending the old password by e-mail;
		\item sending a new password by e-mail;
		\item sending a new temporary password by e-mail;
		\item sending an HTTP link by e-mail (that will allow the requesting user to choose a new password);
		\item sending an HTTPS link by e-mail (that will allow the requesting user to choose a new password).
	\end{itemize}
	The vulnerabilities of this mechanism are detailed in the following sections.
	
	\subsubsection{Other password recovery mechanisms.}
	Many alternatives have been proposed to manage the password recovery mechanisms in a safer way. Phone numbers can be used for user authentication: the website can send users an SMS message or use an automated voice system to call them and provide an account recovery code. As for e-mail-based authentication, this alternative relies on the fact that only the device holder will be able to access a secret that has been sent to the device~\cite{reeder2011password}. Brainard et al., in~\cite{brainard2006fourth}, rely on \textit{trustees}, users previously appointed by her account holder to verify her identity. These people will be contacted in case of emergency (the password is either lost or compromised) and they will be asked to perform the recognition of the requesting user. If the recognition phase is successful they will provide the requesting user with a recovery code, useful to authenticate her on the website.
	
	\section{Adversary Model}
	Given the websites ecosystem, we consider different categories of adversaries. An adversary may be \textit{undetectable} or \textit{detectable}. An undetectable adversary is able to impersonate the user on websites without the victim being aware of it. Conversely, a detectable adversary impersonating the user has a chance of getting the user being aware, or a least suspicious, of the fact that an impersonation happened (or could have happened). An adversary may also be \textit{active} or \textit{passive}. An active attacker interacts with websites in order to get information about the victim (e.g., she can start the recovery procedure on behalf of the victim), while a passive attacker aims to obtain the target user's sensitive information without any interaction with the website. Eventually, both active and passive attackers can use the obtained information in order to access websites by pretending to be the victim. Remembering these distinctions (summarized in~\autoref{tab:attackers_description}), we introduce four different possible attacks against a single target user, taking into account a single target website:
	\begin{itemize}
		\item Mail service provider-level attack;
		\item Web server intruder attack;
		\item Client intruder attack;
		\item Sniffing attack.
	\end{itemize}
	
	\begin{table}[h]
		\setlength{\extrarowheight}{0.1cm}
		\caption{Attackers Types}
		\label{tab:attackers_description}
		\centering
		\begin{tabular}{C{3cm}|C{4cm}C{4cm}}
			\textbf{Type} & \textbf{Detectable} & \textbf{Undetectable} \\
			\hline
			\textbf{Active} & Can interact with the website on behalf of the victim by revealing her existence  & Can interact with the website on behalf of the victim by remaining transparent to her \\
			\hline
			\textbf{Passive} & Cannot interact with the website, but her actions will eventually make the victim aware of her existence  & Cannot interact with the website and will remain transparent to the victim\\
			\hline
		\end{tabular}
	\end{table}
	
	\begin{table}[h]
		\setlength{\extrarowheight}{0.1cm}
		\caption{Attackers Capabilities}
		\label{tab:attackers_capabilities}
		\centering
		\begin{tabular}{C{3cm}|C{3cm}C{3cm}C{3cm}}
			\textbf{Attackers/Access} & \textbf{User e-mails} & \textbf{Website password DB} & \textbf{Website password recovery method} \\
			\hline
			Mail service provider attacker & \cmark & \xmark & \cmark \\
			\hline
			Web server intruder attacker & \xmark & \cmark & \cmark \\
			\hline
			Client intruder attacker & \cmark & \xmark & \cmark \\
			\hline
			Sniffing attacker & \xmark & \xmark & \cmark \\
			\hline
		\end{tabular}
	\end{table}
	
	An \textit{e-mail service provider-level attack} can be undertaken by both a malicious service provider, or a malicious user that has compromised the e-mail service provider. This is the most dangerous attack, as the attacker has access to all the user's e-mails and can obtain a lot of sensitive information. In the \textit{web server intruder attack}, we suppose the attacker was able to violate a website where the user has registered a personal account. So, the adversary can interact with the database that stores all the passwords of registered users. A \textit{client intruder attack} can be undertaken by an adversary that either has violated a user's device by obtaining a remote access, or has stolen it. In the \textit{sniffing attack} we suppose that the attacker has no knowledge about the user, she has neither access to the website's password databases nor to the user's devices. The only information she has, is about the password recovery methods of the website (to obtain this information she could register an account on it and start the recovery procedure). The sniffing adversary can sniff the packets transmitted during the communication between the client and the website. We assume without loss of generality that the sniffing attacker is not able to read the content of the exchanged e-mails (for instance, because the HTTPS protocol is used). The capabilities of attackers are summarized in~\autoref{tab:attackers_capabilities}.
	
	\section{Methodology and Results} \label{Sec:Methodology_Results}
	In this section we provide the methodology employed to analyze the websites' password management security. The methodology involves two independent choices:
	\begin{enumerate}
		\item which website to take into account for the analysis; and,
		\item how to analyze the so chosen websites.
	\end{enumerate}
	
	As for the first choice, we decided to make use of the Amazon Alexa Top Sites web service~\cite{amazon_alexa}. Amazon Alexa Top Sites is a web service that provides a list of websites, ordered by Alexa Traffic Rank. This ranking is determined by a combined measure of both websites' unique visitors and websites' page views, and is updated daily~\cite{amazon_alexa_traffic_ranking}. So, we have selected a subset of countries, respectively England, France, Germany, Italy, and Spain according to their subjectiveness to the GDPR regulation---till Brexit happens, UK is subject to GDPR as well. For each of these countries, we considered the first top 200 websites according to Alexa's ranking. The choice of this parameter is due to the fact that as the ranking goes down, the websites become less and less used--hence, with a reduced impact. 
	We have therefore created an account and started obtaining the websites' URLs we needed. Amazon Alexa provides several ways for obtaining websites information; in particular, the top websites can be selected according to three major divisions: global; by country; and by category. The global division, as the name suggests, shows the most visited websites globally. The division by country, instead, allows to view the most visited websites from specific countries, not necessarily with the domain registered in the country taken into account (e.g., the website \textit{youtube.com} is in the top five of all the countries considered in this analysis, but its domain is registered in the US). The third division is the more generic and allows to select the top 200 websites according to different sub-categories (e.g., adult, arts, computers, recreation, regional, sports, and so on). By selecting the regional category, we are able to choose the continent first and then the specific nations we would like to consider. In this case, Alexa shows the top 200 websites for the countries selected, that represents the top 200 websites with the domains registered in those countries. 
	Once obtained the websites URLs, our goal was to analyze them in order to get information about the password storage management. In the second phase we started by selecting, among the top 200 websites for each nation, the ones that required a user registration. A further filtering step was to remove those websites that, in the user registration phase, required too much information (e.g., bank websites, universities websites, wholesale reseller websites). 
	In detail, of these 1,000 websites, 722 did not require authentication---and hence were excluded by our study---, while out of the remaining 278, we focused on 174---since 104 demanded a complex registration procedure. 
	The results of this filtering are shown in~\autoref{tab:analyzed_websites}.
	
	\begin{table}[h]
		\setlength{\extrarowheight}{0.1cm}
		\caption{Number of analyzed websites (among the top 200 per country)}
		\label{tab:analyzed_websites}
		\centering
		\begin{tabular}{C{2cm}C{3cm}}
			\textbf{Country} & \textbf{Websites (\#)} \\
			\hline
			England & 71\\
			\hline
			France & 31\\
			\hline
			Germany & 19\\
			\hline
			Italy & 36\\
			\hline
			Spain & 17\\
			\hline
			\textbf{Total} & \textbf{174}\\
		\end{tabular}
	\end{table}
	
	\noindent Note that only a portion of the websites taken into consideration offer users the ability to register an account. By considering the number of websites, we decided to manually perform the analysis of each one, in order to obtain more accurate and detailed information---while this activity took quite some amount of time (around three months), the quality of results is unpaired. In particular, for each website we carried out the following steps: 
	\begin{itemize}
		\item we registered an user account;
		\item we pretended to have lost the password and we started the recovery procedure; and,
		\item we collected password recovery information.
	\end{itemize}
	
	\begin{table*}[h]
		\setlength{\extrarowheight}{0.1cm}
		\caption{Websites password recovery mechanisms}
		\label{tab:analysis_result}
		\centering
		\begin{tabular}{C{1.5cm}C{1.5cm}|C{1.75cm}C{1.75cm}C{1.75cm}C{1.75cm}C{1.75cm}}
			\textbf{Country} & \textbf{Websites (\#)}  & \textbf{Old Pw}  & \textbf{New Pw} & \textbf{Temp Pw} & \textbf{HTTP link} & \textbf{\% of vulnerable websites} \\
			\hline
			England & 71 & 1 & 5 & 2 & 16 & 33.8 \\
			\hline
			France & 31 & 0 & 10 & 0 & 7 & 54.84\\
			\hline
			Germany & 19 & 1 & 1 & 0 & 5 & 36.84\\
			\hline
			Italy & 36 & 4 & 11 & 1 & 4 & 55.55\\
			\hline
			Spain & 17 & 2 & 5 & 0 & 1 & 47.06\\
			\hline
			\textbf{Total} & \textbf{174} & \textbf{8} & \textbf{32} & \textbf{3} & \textbf{33} & \textbf{43.68}\\
		\end{tabular}
	\end{table*}
	
	We registered for the websites by using a Gmail account created for the experiment, i.e., \textit{gdpr.experiment@gmail.com}. The results of the analysis are shown in~\autoref{tab:analysis_result}. Each column of the table represents a different method of password recovery adopted by the analyzed websites---these password recovery methods can be observed from left to right in decreasing order of vulnerability. 
	
	\subsection{Password recovery methods}
	In the following, we describe each password recovery method encountered in the analyzed websites. 
	
	\subsubsection{Old Password.}
	The most vulnerable websites are the ones that use the \textit{Old Pw} method for the password recovery phase. In detail, after having completed the ``forgotten password'' procedure, the website sends the original password of the user in her registration e-mail address. Hence, it appears that websites do not make use of hash functions or other mechanisms in order to avoid to store the passwords in cleartext. Considering that most of the users make use of the same password to access many different services~\cite{reusing_password}, obtaining access to the database of password stored by the website (by an hacker attack or an internal website error) would seriously jeopardize both the security and the privacy of all the registered users. This method of storing password is deprecated since at least 30 years~\cite{fighting_computer_crime}.
	
	\subsubsection{New Password.}
	In this recovery method, after the ``forgotten password'' procedure, the website sends a new password to the registration e-mail address of the user, without obliging her to change the password after the first access. The security of passwords sent by using this method is summarized in~\autoref{tab:password_analysis}. In this analysis we take a lenient stance and consider respectively: \textit{strong} the passwords with at least $2^{70}$ possible combinations; \textit{medium} the passwords with at least $2^{50}$ possible combinations; and \textit{weak} all the others. We have obtained this data by requesting a new password 125 times, and by analyzing the type of provided password. For instance, \textit{VUZUK8G3}, \textit{484ad5b5}, \textit{ugrxpn} are three of the passwords provided by a given website following the recovery procedure. It is possible to see that all the passwords use only uppercase or lowercase letters, sometimes numbers, never special characters, with an overall length of 8. In this case, the strength has been computed as $62^8 \approx 2^{48}$ . 
	
	\begin{table*}[h]
		\setlength{\extrarowheight}{0.1cm}
		\caption{Password robustness analysis}
		\label{tab:password_analysis}
		\centering
		\begin{tabular}{C{2cm}|C{1.30cm}|C{1.30cm}|C{1.30cm}|C{1.30cm}|C{1.30cm}|C{1.30cm}|C{1.30cm}}
			\hline
			\textbf{} & $>2^{10}$ & $>2^{20}$ & $>2^{30}$ & $>2^{40}$ & $>2^{50}$ & $>2^{60}$ & $>2^{70}$ \\
			\hline
			New & 100\% & 93.75\% & 75\% & 62.5\% & 3.125\% & 3.125\% & 3.125\% \\
			\hline
			Temporary & 100\% & 100\% & 100\% & 100\% & 33.33\% & 0 & 0 \\
			\hline
		\end{tabular}
	\end{table*}
	
	\noindent Note that only 3.125\% of the new passwords provided to the users by the website have what could be considered a decent level of security, while more than 90\% are considered weak. This password choice makes the websites vulnerable to brute-force attacks, where malicious users can try to guess the users' password after requesting the re-sending on behalf of the victims. 
	
	\subsubsection{Temporary Password.}
	The \textit{Temp Pw} method consists in sending a temporary password to the requesting user, where this password must be changed on her next access. Only few websites make use of this recovery method, and those who do, they send either weak or medium security password to the requester e-mail account~(see \autoref{tab:password_analysis}).
	
	\subsubsection{HTTP link.}
	Almost 19\% of the websites use the \textit{HTTP link} as recovery method. In this case, once the user has requested for the password recovery, the web service sends an HTTP link to her registration e-mail address. By clicking on the link, the requester is redirected to the website on which she can enter a new password that will be associated with her account. We consider vulnerable those websites that use this recovery method. Indeed it is well-known that the HTTP protocol does not provide insurance with respect to attacks such as man in the middle~\cite{man_in_the_middle_attack}, or even simple snooping.
	Considering that all communications between user browser and websites are not encrypted, a malicious user can intercept the message exchange, eavesdrop and modify the communication, compromising both its confidentiality and integrity. 
	
	\subsubsection{HTTPS link.}
	The last analyzed recovery method of the websites is the \textit{HTTPS link} one, in which the link sent to the user to let her choose a new password is based on HTTPS protocol. This method is safer with respect to the others but it is subject to being exploited as well, as we will detail in the following.
	
	\subsection{Attackers capabilities}
	In this section we describe the capabilities of attackers as well as their characteristics. Results of this study are summarized in~\autoref{tab:passive_attackers_synoptic_tabel} for passive attackers, and in~\autoref{tab:active_attackers_synoptic_tabel} for active attackers.
	
	\subsubsection{Mail service provider-level attacker.} 
	A \textit{passive e-mail service provider-level attacker} may obtain the user's log-in information in any case, regardless of the password recovery methods adopted by the website. In fact, this kind of attacker has access to the e-mails of the victim, the emergency authentication mechanisms currently adopted by websites. In case of \textit{Old Pw} or \textit{New Pw} recovery methods, the attacker can even remain undetectable. In fact, by easily reading the password inside the e-mail, the attacker and the victim would share the same account unbeknownst to the latter. The attacker is forced to adopt a detectable method in the other cases (i.e., the adversary can use either the temporary password or both the HTTP and HTTPS link to log-in with the credentials of the victim, but once logged in she is forced to change the password, by no longer granting access to the victim). In this situation, the provider could obtain the credentials (by reading either the new or the temporary password) as well as the links (either HTTP or HTTPS) and destroy the received e-mail. The user would not receive the e-mail but could attribute the fact to a website malfunction. In the meantime, the adversary may have entered, obtained the information she needed, and logged out. There is a good chance of not being suspected at all. \newline
	An \textit{active e-mail service provider-level attacker} could remain undetectable if the target website stores the passwords of registered users in clear. In fact, by starting the recovery procedure, she can obtain the original password of the user by e-mail. This e-mail will be deleted from the system as soon as possible. In all other cases, the password would change and the e-mail service provider-level attacker will delete the compromising e-mail. The victim would have no longer access to the website but could attribute the fact to either a website malfunction or to having forgotten the password.
	
	\subsubsection{Web server intruder attacker.}
	The information that a \textit{passive web server intruder attacker} can gain is strongly dependent on the password storage management of the website she has violated. Indeed, if passwords of the registered users are stored in clear text (i.e., without the support provided by hash functions), the attacker could transparently make use of the credentials of the victim to access the target website.\\
	The capability does not change from the perspective of an \textit{active web server intruder attacker}, because they would still depend to the storage methods. In addition, note that the victim would be aware of an attack as it would receive e-mails with instructions for recovering the password. 
	
	\subsubsection{Client intruder attacker.}
	A \textit{passive client intruder attacker}, with remote or physical access to the device of the victim, would have most of the time exactly the same capabilities of an e-mail service provider-level attacker (i.e., there is often no need to insert a password to access services, most of users allow the browser to remember it~\cite{password_life_cycle}). Furthermore, either remote or physical access to the device would allow the attacker to find any password file written by the victim~\cite{password_memorability}, so the attacker capabilities depend on the victim's behavior.
	If the victim does not allow the browser to remember the information about her credentials, an attacker with remote or temporarily physical access to the device could install a key-logger software~\cite{gpu_keylogger}, that would capture user input and provides the attacker the information about the keys typed (including the passwords of the websites that the victim will visit in that session). \\
	An \textit{active client intruder attacker} would have the same capabilities of the active e-mail service provider-level attacker, because of the access to the e-mail account of the victim.
	
	\subsubsection{Sniffing attacker.}
	A \textit{passive sniffing attacker} can obtain the access information of a target user mainly in two ways: she can intercept the HTTP communications, as well as undertake a man-in-the-middle attack~\cite{man_in_the_middle_attack} during an HTTPS communication, or she can try to brute-force a target website until the victim's credentials are guessed~\cite{burp_bruteforce}. In this case, unless the website implements some type of protection (e.g., CAPTCHA, blocking of the account log-in for a few seconds after \textit{n} attempts) or warning system (e.g., sending an e-mail to the account owner after a number of failed log-in attempts), the attacker would be undetectable due to her knowledge of the password of the victim. \\
	The capabilities of an \textit{active sniffing attacker}, instead, would be subject to various changes. If the server adopts the old Pw recovery method (e.g., it stores the registered users passwords in clear text and, when requested, it sends back the original user's password), the sniffing attacker can ask for the password recovery on behalf of the user but it would be both detectable (given that the user will receive the e-mail with the old password) and useless (she should brute-force the same password as before). In case the web server adopted either New or Temp Pw password recovery method, instead, the active sniffing attacker would have useful information about the password she has to brute-force. In fact, even if the user has a very strong password, the attacker could apply for a new (or a temp) one on her behalf. In this way, the attacker would know exactly the structure and the level of security of the new website-generated password and could brute-force it accordingly. For example, the target user has a very hard-to-guess password, and the attacker starts the password recovery methods on a target website in her behalf. The attacker, by registering a personal account on that website, knows how temp and new passwords are chosen (e.g., only numbers, both capital letter and numbers, etc.), so she would have useful information about the password structure as well as its length. In case of HTTP or HTTPS links used for password recovery, the attack will become detectable and the attacker would not gain any additional information about the password with respect to the passive attacker ones. 
	
	\begin{table*}[htbp]
		\setlength{\extrarowheight}{0.1cm}
		\caption{Synoptic table related to passive attackers}
		\label{tab:passive_attackers_synoptic_tabel}
		\centering
		\begin{tabular}{C{1.95cm}|C{1.95cm}|C{1.95cm}|C{1.95cm}|C{1.95cm}|C{1.95cm}}
			\hline
			\textbf{Attacks / Recovery methods} & \textbf{Old Pw} & \textbf{New Pw} & \textbf{Temp Pw} & \textbf{HTTP link} & \textbf{HTTPS link} \\
			\hline
			Mail service provider-level & \textit{undetectable} & \textit{undetectable} & \textit{detectable} & \textit{detectable} & \textit{detectable} \\
			\hline
			Web server intruder & \textit{undetectable} & Depends on the storage method & Depends on the storage method & Depends on the storage method & Depends on the storage method \\ 
			\hline
			Client intruder & \textit{undetectable} & \textit{undetectable} & \textit{detectable} / Depends on the user's behavior & \textit{detectable} / Depends on the user's behavior & \textit{detectable} / Depends on the user's behavior \\
			\hline
			Sniffing & \textit{undetectable} & \textit{undetectable} & \textit{undetectable} & \textit{undetectable} & \textit{undetectable} \\
			\hline
		\end{tabular}
	\end{table*}
	
	Note that two of the main differences between passive and active attackers involve attack timing and attack extension. On the one hand, active attackers may get information about the victim's credentials at any time, while passive ones must wait for a move from the user. On the other hand, an active attacker must take immediate actions while a passive attacker can be implemented as a fragment of autonomous software that will be triggered by certain events (e.g., starting of a recovery password procedure by the user, receiving a mail with credentials/links). This means that passive attacker softwares could be easily and quickly propagated.
	
	\begin{table*}[htbp]
		\setlength{\extrarowheight}{0.1cm}
		\caption{Synoptic table related to active attackers}
		\label{tab:active_attackers_synoptic_tabel}
		\centering
		\begin{tabular}{C{1.95cm}|C{1.95cm}|C{1.95cm}|C{1.95cm}|C{1.95cm}|C{1.95cm}}
			\hline
			\textbf{Attacks / Recovery methods} & \textbf{Old Pw} & \textbf{New Pw} & \textbf{Temp Pw} & \textbf{HTTP link} & \textbf{HTTPS link} \\
			\hline
			Mail service provider-level & \textit{undetectable} & \textit{detectable} & \textit{detectable} & \textit{detectable} & \textit{detectable} \\
			\hline
			Web server intruder & \textit{undetectable} & \textit{detectable} / Depends on the storage method & \textit{detectable} / Depends on the storage method & \textit{detectable} / Depends on the storage method & \textit{detectable} / Depends on the storage method \\ 
			\hline
			Client intruder & \textit{undetectable} & \textit{detectable} & \textit{detectable} & \textit{detectable} & \textit{detectable} \\
			\hline
			Sniffing & \textit{detectable - useless} & \textit{detectable - simpler} & \textit{detectable - simpler} & \textit{detectable} & \textit{detectable} \\
			\hline
		\end{tabular}
	\end{table*}
	
	\section{Countermeasures}
	In this section we describe some countermeasures that can be taken into account to mitigate the different types of attacks exposed in previous sections.
	
	\subsubsection{Good Practices.}
	Basic security guidelines dictate, since at the last 40 years, that passwords should never be stored in clear text. For instance, the password could be hashed by using a dash of salt, different for each password. The salts could be stored on the same server of the passwords, because the mere possession of the salt would not help the attacker to find out the passwords of the users (i.e., it is very hard to find salted hashes in rainbow tables unless they were prohibitively large). Another solution can be the implementation of slow hashes by websites. The website would suffer negligible delay every time a user registers an account or logs in, but the password table attack by the attacker would be infeasible due to the time required. SHA-1, SHA-256, and MD5 are relatively fast hash functions, while bcrypt, PBKDF2, and scrypt could be used for this purpose. 
	
	Moreover, the password length plays a fundamental role. Computers, Field-programmable gate array (FPGA), and Application-specific integrated circuit (ASIC) technologies, are becoming fast and able to brute-force both non-salted and salted passwords. 8-character passwords are not enough robust even by using a combination of numbers, lowercase letters, uppercase letters, and special characters. In fact, even by considering a random password the complexity will be equal to $95^{8}$ ($< 2^{56}$), a negligible number if we consider that the validation of a block of BitCoin transactions requires, as of writing, $2^{72}$ hashes~\cite{blockchain_last_block}.
	
	Since currently, for most websites, the ability to access an e-mail address identifies the user as the owner of that address, some further good practices could be listed as follows:
	\begin{itemize}
		\item do not allow the browsers/websites (especially the e-mail service providers' ones) to save the password. Better to waste a few seconds to rewrite it than losing control of your identity on the web;
		\item always log-out and remove open sessions;
		\item before inserting either passwords or other sensitive information, make sure that there are no key loggers in your system (at least the software ones, more easily detectable);
		
	\end{itemize}

	\subsubsection{Cross-Mail service providers secret distribution.}
	In order to avoid e-mail service provider level attacks, it could be useful to rely on more than one e-mail service providers, basically by not trusting any of them. When the user wants to register an account on a website, she will provide it with more than one e-mail address, each belonging to a different e-mail service provider. Once the user has lost her password (or someone has stolen it), she can start the password recovery procedure. The website will generate a new password and will split it into a number of chunks equal to the number of different e-mails account provided by the user during the registration. Finally, the website will send each fragment to a different email address. Once the user has received all the fragments, she can put them together in order to rebuilt the original website-crafted password. A malicious e-mail service provider could only have a partial view of the whole password and, starting from that, it will be unfeasible to rebuilt the original one. The user should be careful not to use all the accounts on the same device, otherwise in case of either client intruder attack or loss of the device, the attacker would have access to all the victim's e-mail addresses. This countermeasure allows to protect from e-mail service provider attacks by sacrificing a little the usability.
	
	\subsubsection{*-based password request.}
	Previous countermeasures, unless particular precautions are taken by the users, are still subject to the client intruder attacks. In order to mitigate this issue, these solutions can be integrated with the *-based password request mechanism. The *-based password request mechanism requires the user to enter her e-mail account password whenever certain events occur. Several variations of the mechanism can be used, for example: a time-based password request mechanism requires the user to enter the e-mail account password when a pre-determined amount of time has passed (e.g., if the user wants to open either a new or an old e-mail after 30 minutes she entered the password the last time, she will be prompted for the password again); event-based password request mechanism requires the user to enter the password when a pre-determined sequence of events has occurred (e.g., after n e-mails are opened by the user, the password will be requested again). There is the possibility of integrating multiple variations of the highlighted mechanisms to provide the user with a higher degree of security (e.g., time-based password request mechanism used in conjunction with the event-based password request mechanism: both mechanisms are considered to decide whether and when to request the password to the user). The parameters of these mechanisms can also be learnt automatically by the system after an initial training phase, to provide the user with a tailored experience.
	
	By using this mechanism, an attacker who has a remote access to the victim's system (or a physical access, because in possession of the victim's device), would have a limited access to the e-mails, as she would be required to enter the password (which she does not know).
	
	\subsubsection{Choosing a good service provider.}
	There are a number of services that have been designed to preserve both the privacy and improve the safety of users. Speaking of e-mail service, the most used are ProtonMail, Tutanota, and ShazzleMail.
	ProtonMail~\cite{protonmail}, as well as Tutanota~\cite{tutanota}, they make use of end-to-end encryption to hide the content of the e-mails to everyone (including the mail service provider itself) except for the interlocutors. ShazzleMail~\cite{shazzlemail} is a free private e-mail application that turns smart phones into e-mail servers, and users can deliver messages directly to the receiver by using an SSL encrypted channel with no server copies.
	
	\section{Conclusion}
	In this paper, we first provided a survey of both user authentication mechanisms implemented by websites, and password recovery mechanisms currently adopted. Then, we modeled an attacker with different capabilities and we showed how simply she is able to get the users' confidential information---e.g. recovering the password, or accessing e-mail. 
	Later, we performed a detailed analysis of users password management for Alexa's top 200 websites (see Section \ref{Sec:Methodology_Results}) of five European countries, respectively England, France, Germany, Italy, and Spain. It is worth noting that analyzed websites are among top ones according to the Alexa's ranking, so, they are supposed to be way too secure compared to those present in the lower ranking.
	Our results show that, nowadays, almost 44\% of the analyzed websites are vulnerable---simply because lacking adoption of  best practices, and hence subject to the fines complied by the GDPR.
	We are currently working on a P2P system based on Blockchain technology to manage both authentication and password recovery of websites.
	
	\bibliographystyle{splncs04}

\end{document}